\documentclass{article}

\usepackage[english]{babel}
\usepackage{geometry}
\usepackage{graphicx}
\usepackage{cite}
\usepackage{amssymb,amsmath}
\usepackage{pstricks}
\usepackage{amsfonts}

\def\BM{\begin{displaymath}}
\def\EM{\end{displaymath}}
\def\BE{\begin{equation}}
\def\EE{\end{equation}}
\def\BA{\begin{eqnarray}}
\def\EA{\end{eqnarray}}
\def\BF{\begin{figure}}
\def\EF{\end{figure}}
\def\BC{\begin{center}}
\def\EC{\end{center}}
\let\TW=\textwidth
\def\sfrac#1#2{\mbox{$\frac{#1}{#2}$}}
\def\DDt{\sfrac{\rm d}{{\rm d}t}}
\def\RG{\ensuremath{R_{\rm g}}}
\def\RT{\ensuremath{R_{\rm t}}}

\title{Pattern fluctuations in transitional plane Couette Flow}

\author{Joran Rolland, Paul Manneville \\
{\small Laboratoire d'Hydrodynamique de l'\'Ecole Polytechnique, 91128 Palaiseau, France}}

\date{\today}
\begin{document}
\maketitle

\begin{abstract}

In wide enough systems, plane Couette flow, the flow established between two parallel plates translating in opposite directions, displays alternatively
turbulent and laminar oblique bands in a given range of Reynolds
numbers $R$. We show that in periodic domains that contain a few
bands, for given values of $R$ and size, the orientation and the
wavelength of this pattern can fluctuate in time. A procedure is
defined to detect well-oriented episodes and to determine the statistics
of their lifetimes. The latter turn out to be distributed according
to exponentially decreasing laws. This statistics is interpreted in
terms of an activated process described by a Langevin equation whose
deterministic part is a standard Landau model for two interacting
complex amplitudes whereas the noise arises from the turbulent background.

\end{abstract}

\section{Introduction\label{s1}}

The main features of the transition to turbulence are well understood in systems prone to a linear instability like convection where chaos emerges at the end of an instability cascade. A much wilder transition is observed in wall-bounded shear flows for which the laminar and turbulent regimes are both possible states at intermediate values of the Reynolds number $R$, the natural control parameter, whereas no linear instability mechanism is effective. A direct transition can take then place via the coexistence of laminar and turbulent domains in physical space. Two emblematic cases are the pipe flow and plane Couette flow (PCF), the simple shear flow developing between two parallel plates translating in opposite directions. Both of them are stable against infinitesimal perturbations for all values of $R$ and become turbulent only provided sufficiently strong perturbations are present. In both cases, strong hysteresis is observed and, upon decreasing $R$, the turbulent state can be maintained down to a value $R_{\rm g}$. Above $R_{\rm g}$, turbulence remains localised in space, in the form of turbulent puffs in pipe flow and turbulent patches in  PCF. A striking property of PCF or counter-rotating cylindrical Couette flow (CCF) is the
spatial organisation of turbulence in alternatively turbulent and laminar
oblique bands that takes place in large enough systems in a specific
range of Reynolds numbers \cite[Ch.7]{Ma10}.
This regime was studied in depth
at Saclay by Prigent {\it et al.} \cite{Petal03}.
It can be obtained by decreasing
the Reynolds number continuously from featureless turbulence below
\RT, the Reynolds number above which the flow is uniformly
turbulent, or triggered from laminar flow by finite amplitude perturbations 
above \RG, the Reynolds number below which laminar flow is expected
to prevail in the long time limit. A similar situation is observed in pipe flow but things are complicated by the global downstream advection so that the existence of a threshold $R_{\rm t}$ above which turbulence is uniform is still a debated matter. In contrast for PCF, the pattern is essentially
time-independent and can be characterised by two wavelengths $\lambda_x$
and $\lambda_z$ in the streamwise and spanwise direction, $x$ and $z$
respectively,%
\footnote{In the case of CCF, the pattern is time-independent in a frame
that rotates at the mean angular velocity and the axial (azimuthal)
direction corresponds to the spanwise (streamwise) direction.}
or equivalently by a wavevector $\mathbf{k}=(k_x,k_z)$ with
$k_{x,z}=2\pi/\lambda_{x,z}$. From symmetry considerations, two
orientations are possible, corresponding to two possible combinations
$(k_x,\pm k_z)$. Whereas a single orientation is present sufficiently
far from \RT\ so that either mode $(k_x,+k_z)$ or mode
$(k_x,-k_z)$ is selected, patches of one or the other orientation have
been reported to fluctuate in space and time when $R$ approaches
\RT\ from below \cite[Figs.~2 \& 3]{Petal03}.
The main features of the bifurcation diagram
could then be accounted for at a phenomenological level by an approach
in terms of Ginzburg--Landau equations subjected to random noise
featuring the small-scale turbulent background.

This patterning was reproduced by Duguet {\it et al.} \cite{Detal10}
using fully resolved numerical simulations in an extended system of size comparable with that of the Saclay apparatus but the computational load
was so heavy that a statistical study of the upper transitional range was inconceivable. Earlier, Barkley \& Tuckerman \cite{BT05} also succeeded
in obtaining the bands by means of fully resolved simulations with less
computational burden but using narrow elongated domains aligned with
the pattern's wavevector.
By construction, the fluctuating domain regime could not be obtained,
whereas a re-entrant
featureless turbulence regime, called `intermittent' was obtained closer
to \RT.

In our previous work on this problem, we first showed that full
numerical resolution was not necessary to obtain realistic patterning
but that a good account of the long range streamwise correlation of
velocity fluctuations was essential \cite{MRtcfd}. This next incited us
to consider reduced-resolution simulations in systems of sizes sufficient
to contain at least an elementary cell $(\lambda_x,\lambda_z)$ of the
pattern \cite{RMepjb}, thus avoiding the orientation constraint inherent
in the Barkley--Tuckerman approach. Here, we expand our previous work
to focus on pattern fluctuations in the upper part of the PCF's bifurcation
diagram when $R$ approaches \RT\ from below, taking the best possible use
of the inescapable resolution lowering to perform long duration
simulations, so as to obtain meaningful statistics about the dynamics of
this regime.

Systems considered in our numerical experiment, to be described in
\S\ref{s2}, produce patterns with a few wavelengths.
In the neighbourhood of \RT,
fluctuations manifest themselves as orientation changes in time instead
of the spatiotemporal evolution of well-ordered patches. It turns out that
episodes of well-formed pattern between two orientation changes can be
identified reliably, so that the lifetimes of such episodes can be
measured and their average determined as a function
of $R$. The Langevin approach initiated by Prigent {\it et al.}
in~\cite{Petal03} was resumed in \cite{RMepjb} as providing an
appropriate framework to interpret our numerical results.
Orientation fluctuations were taken into account but their detailed
statistical properties left aside, which are the subject of the present
paper.

In the context of pattern formation, the Langevin/Fokker--Planck
approach has a long history, dating back to the 1970's when it was
applied to convecting systems \cite{Gr74}. Noise of thermal
origin is however extremely weak so that the region of parameter space
where the system is sensitive to this noise is exceedingly narrow
\cite{HS92} and nontrivial effects can be observed only in very specific
conditions \cite{Setal00}. When applying the approach to the description
of the bifurcation from featureless turbulence to pattern in shear flows,
Prigent {\it et al.} \cite{Petal03} implicitly took for granted
that the noise intensity was an adjustable parameter linked to the turbulent background at $R>\RT$. Here we extend the analysis started in
\cite{RMepjb} within this conceptual framework, the subject of \S\ref{s3},
and analyse simulation results presented in \S\ref{s4} in the light of this
theory. We conclude in \S\ref{s5} by discussing how well this approach
is suited to describe mode competition and intermittent re-entrance of
featureless turbulence \cite{BT05,RMepjb} and, more generally, how the
noisy temporal dynamics of coherent modes can hint at the
spatio-temporal nature of transitional wall-bounded flows and explain
the exponentially decreasing probability distributions of residence
times or decay times often observed in this field \cite{Eetal08}.

\section{Conditions of the numerical experiment\label{s2}}

\subsection{Numerical procedure\label{s2.1}}

Direct numerical simulation (DNS) of the incompressible Navier--Stokes equations in the geometry of PCF are performed using Gibson's open source
code {\sc ChannelFlow} \cite{refgibs} that assumes no-slip boundary conditions at the plates driving the flow and in-plane periodic boundary
conditions. The parallel plates producing the shear are placed
at a distance $2h$ from each other in the wall-normal direction $y$, they
move at speeds $\pm U$ in the streamwise direction $x$, $z$ labelling the
spanwise direction. The length unit is $h$, the velocity unit $U$, the time 
unit $h/U$, and the Reynolds number $R=Uh/\nu$, where $\nu$ is the
kinematic viscosity of the fluid. The problem is completely specified when 
the in-plane dimensions $L_x$ and $L_z$ of the set-up are chosen.
The perturbation to the laminar flow $\mathbf U=y\,\mathbf{\hat x}$ is
noted $\mathbf{u}$, so that $\mathbf{u}^2$ is the local Euclidian distance
to the base flow squared.
Periodic in-plane boundary conditions allow the definition of the
wave-vectors
$k_{x,z}=2\pi n_{x,z}/L_{x,z}$, where the wavenumbers $n_{x,z}$ are
integers. Without loss of generality, we can assume $n_x\ge 0$.

The resolution of the simulation is fixed by the number $N_y$ of Chebyshev
polynomials used to represent the wall-normal dependence, and the numbers
$N_{x,z}$ of collocation points used to evaluate the nonlinear terms in pseudo-spectral scheme of integration of the Navier--Stokes
equations. The number of Fourier modes involved in the simulation is then
$2N_{x,z}/3$, owing to the 3/2-rule applied to de-aliase the velocity
field. The computational load necessary to obtain meaningful results
in sufficiently wide domains with fully resolved simulations is unrealistically heavy. Accordingly, we take advantage of our
previous work devoted to the validation of {\it systematic
under-resolution\/}
as a {\it modelling strategy}~\cite{MRtcfd}. In that work, we showed that
qualitatively excellent and quantitatively acceptable results could be
obtained by taking $N_y=15$ and $N_{x,z}=8L_{x,z}/3$. The price to be
paid for the resolution lowering was apparently just a downward shift
of the range $[\RG,\,\RT]$ in which the bands are obtained,
but everything else was preserved, including wavelengths. Of course,
as far as resolution is concerned, the finest is the best on a strictly
quantitative basis but we do not expect that the observed trends and
our conclusions be sensitive to our rules to fix $N_y$ and $N_{x,z}$.

\subsection{Orientation fluctuations.}

In this article we consider domains able to contain pattern with one or
two elementary cells, i.e. $L_{x,z}=|n_{x,z}|\lambda_{x,z}$
where $n_x=1$ or 2 and $n_z=\pm1$ or $\pm2$.
According to \cite{Petal03}, in PCF wavelength $\lambda_x$ is found to be
approximately equal to
$110$ over the whole range $[\RG,\,\RT]$, while wavelength
$\lambda_z$ varies as a function of $R$ in the range $[40,\,85]$
These observations serve us to fix the size of the
systems that we are going to consider below.
As shown in \cite{RMepjb}, the specificity of such systems
is to convert the spatio-temporal evolution of fluctuating domains
observed in the neighbourhood of \RT\  into the temporal evolution
of coherent patterns characterised by the amplitudes of the corresponding
fundamental Fourier modes; possible orientation changes are associated
with changes of sign of the spanwise wavenumbers. Close enough to
\RT, there is also some probability that featureless turbulence,
the state that prevails for $R>\RT$, be observed transiently, which
is akin to the intermittent regime identified in~\cite{BT05}.
In contrast, in the lowest
part of the transitional range, close to \RG, the orientation
remain frozen as expected for well-formed steady oblique bands. We
first illustrate this phenomenon using snapshots of $\mathbf{u}^2$
in Figures~\ref{fig1} and~\ref{fig2}.
\BF
\BC
\includegraphics[width=0.24\TW,clip]{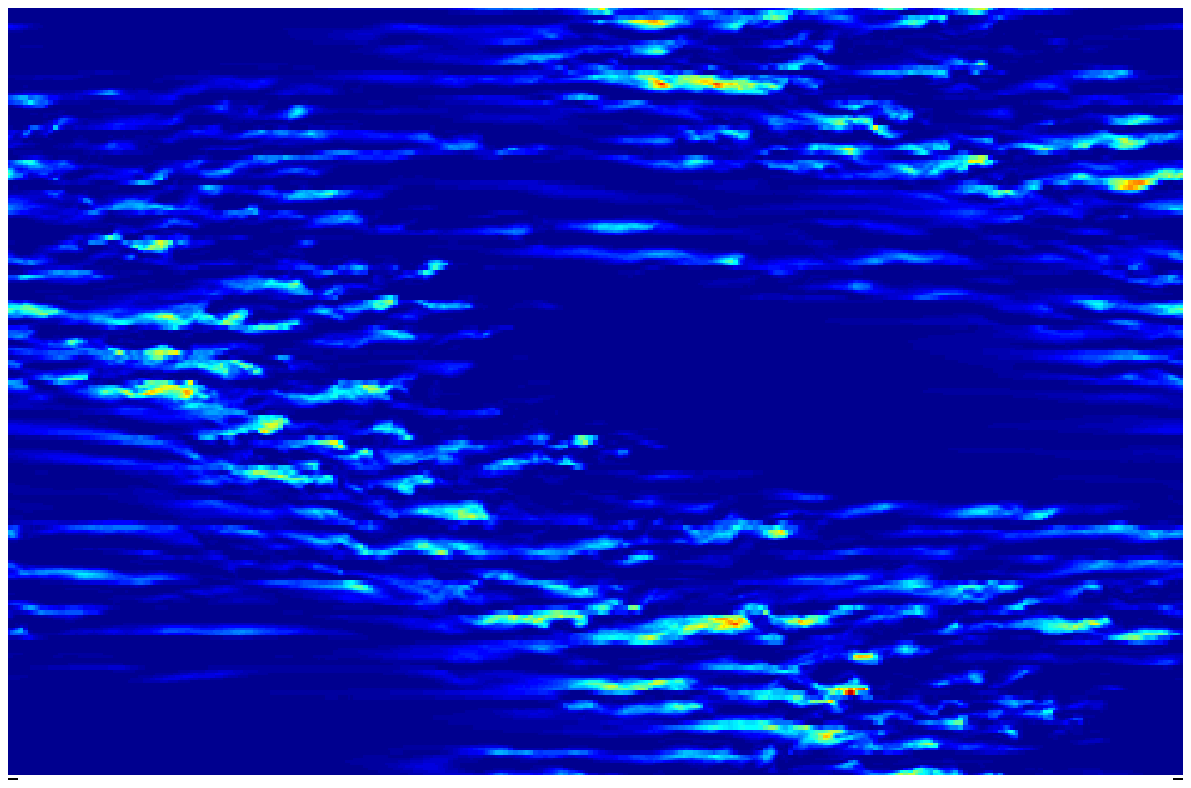}\hfill
\includegraphics[width=0.24\TW,clip]{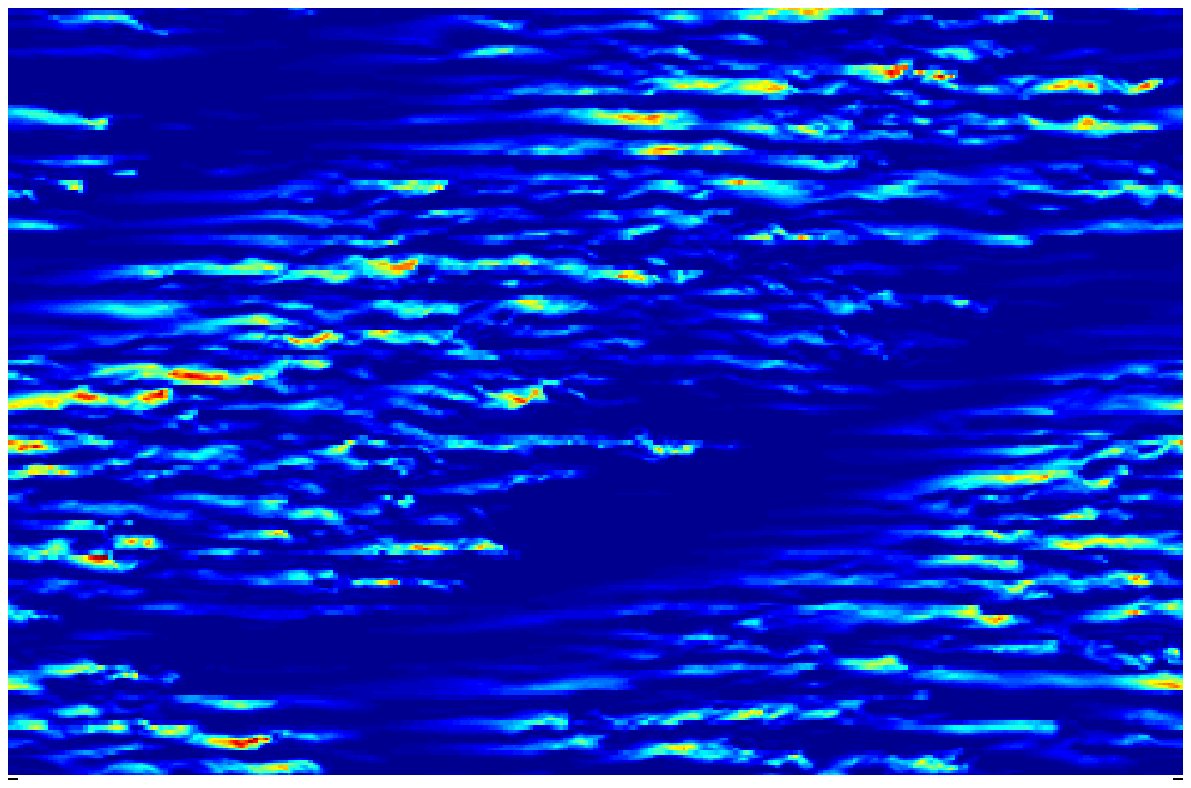}\hfill
\includegraphics[width=0.24\TW,clip]{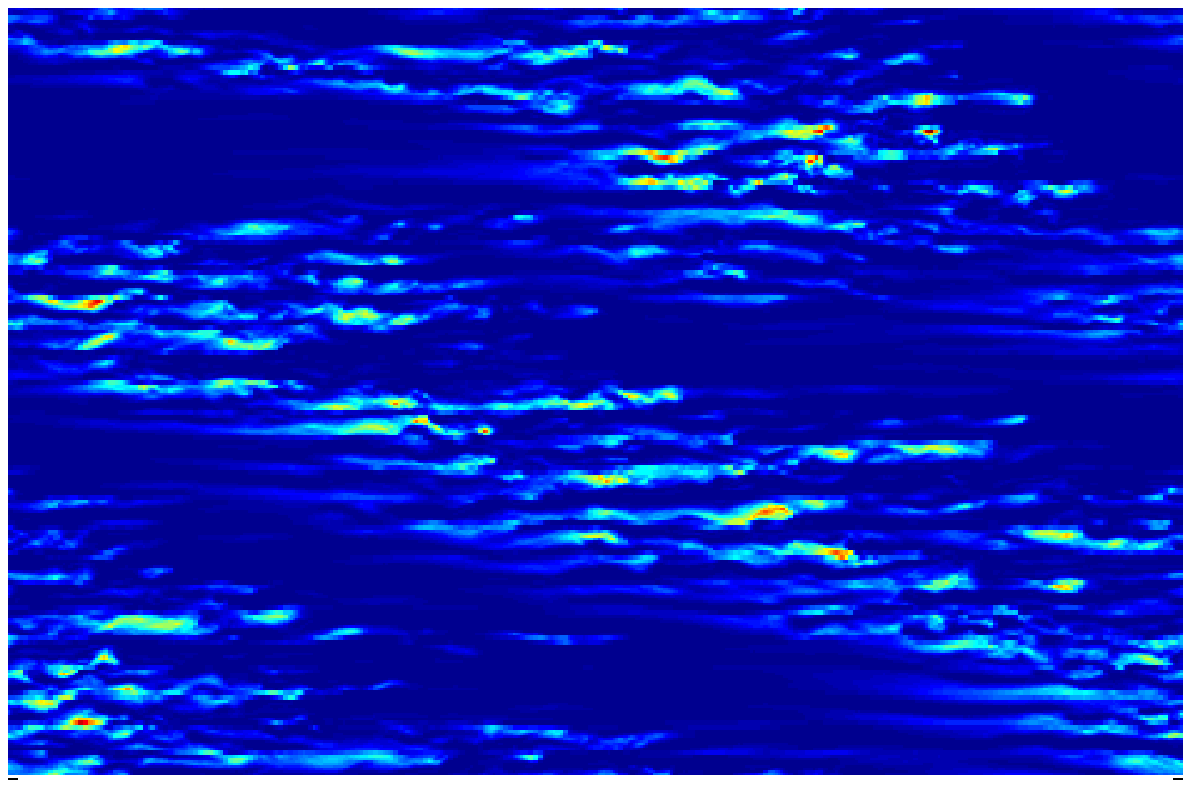}\hfill
\includegraphics[width=0.24\TW,clip]{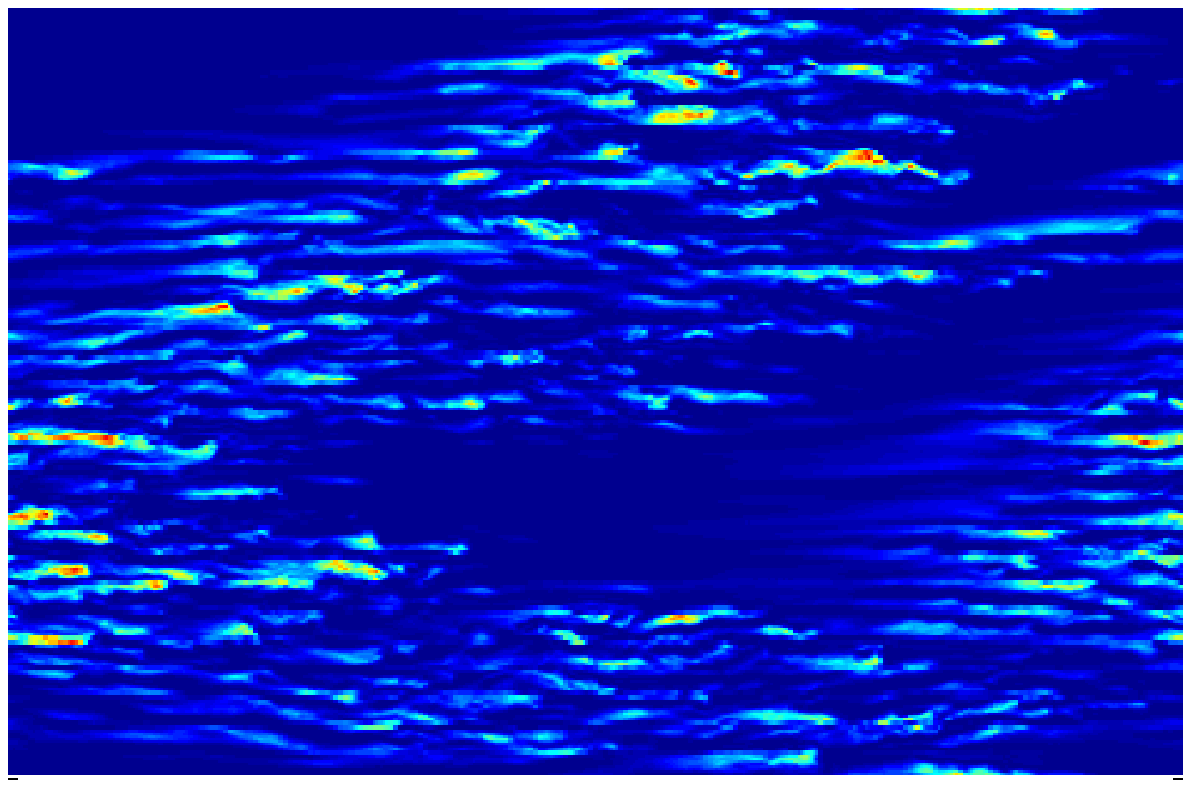}
\EC
\caption{Snapshots of $\mathbf{u}^2$ in the plane $y=-0.57$ for $R=315$
in a system of size $L_x\times L_z=128\times84$. From left to right:
one band pure state with each of the two possible orientations ($n_x=1$, $n_z=+1$) or ($n_x=1$, $n_z=-1$), two band pure state  ($n_x=1$, $n_z=+2$), and mixed or defective pattern. Deep blue corresponds to laminar flow.
\label{fig1}}
\EF
\BF
\BC
\includegraphics[width=0.32\TW,clip]{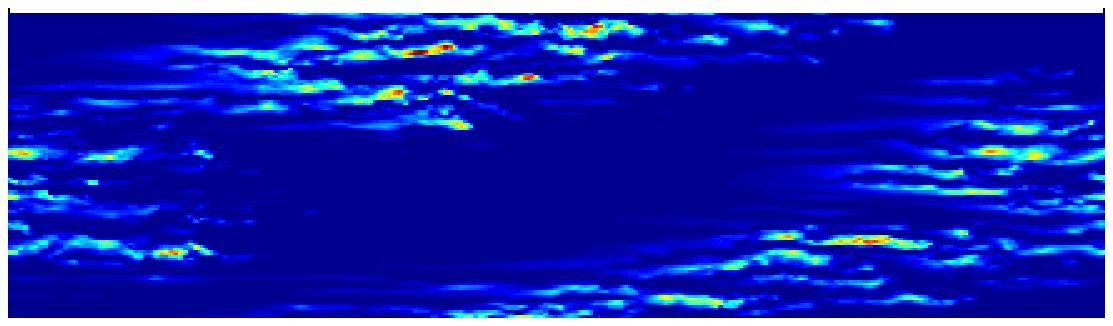}\hfill
\includegraphics[width=0.32\TW,clip]{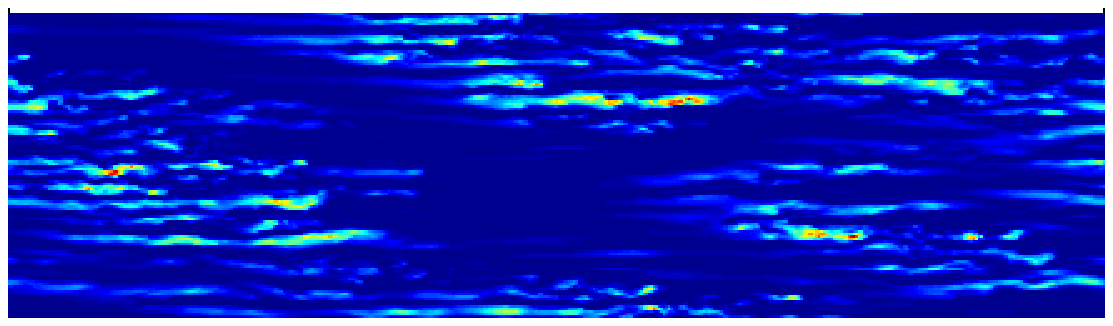}\hfill
\includegraphics[width=0.32\TW,clip]{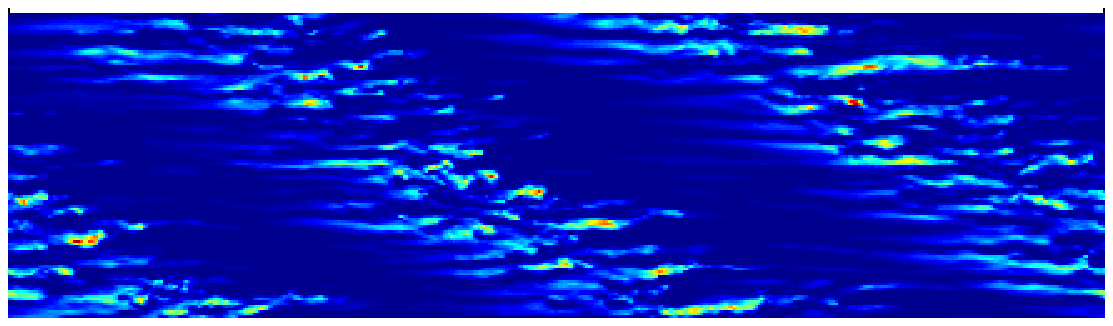}
\EC
\caption{Snapshots of $\mathbf{u}^2$ in the plane $y=-0.57$ for $R=290$
in a system of size  $L_x\times L_z=170\times 48$. From left to right: ($n_x=1$, $n_z=-1$), ($n_x=1$, $n_z=+1$),
and ($n_x=2$, $n_z=+1$). \label{fig2}}
\EF
The left and centre panels of
Fig.~\ref{fig1} display well-oriented patterns or `pure states' showing
the organised cohabitation of laminar and turbulent flow; an example of
defective pattern or `mixed state' without much spatial organisation is
shown in the right panel.
(Orientation defects between well-oriented domains require wider systems
to be clearly identified as such.) Figure~\ref{fig2} similarly displays
snapshots of $\mathbf{u}^2$ obtained in a narrower but longer system.

Typically, during long-lasting simulations at given $L_{x,z}$ and $R$,
the flow displays a pure pattern for some time, then experience a brief
defective stage, and next recovers a pure state, possibly with different
orientation or/and wavelength, and so on.
The spatial organisation of the pattern is detected {\it via\/}
the Fourier transform of the perturbation velocity field
$\hat{\mathbf{u}}$. It turns out that most of the information about
the modulation is encoded in the amplitude of the dominant wavenumber
\cite{Petal03,Betal08,RMepjb}. We consider time series of 
\BE
m^2(n_x,n_z,t)=\frac12\int_{-1}^{+1}|\hat{u}_x(n_x,y,n_z,t)|^2\,{\rm d}y\,,
\label{E-m}
\EE
which thus characterises a flow pattern with wavelengths
$(\lambda_x,\lambda_z)=(L_x/n_x,L_z/|n_z|)$ and orientation given by
the sign of $n_z$. In the present study, we focus on the amplitude of
the turbulence modulation in the flow and not on its phase, i.e. on the
position of the pattern in the system, which was shown to be a random function of time \cite{RMepjb}.

An example of such time series is displayed in Fig.~\ref{fig3}.
\BF
\BC
\includegraphics[width=0.9\TW]{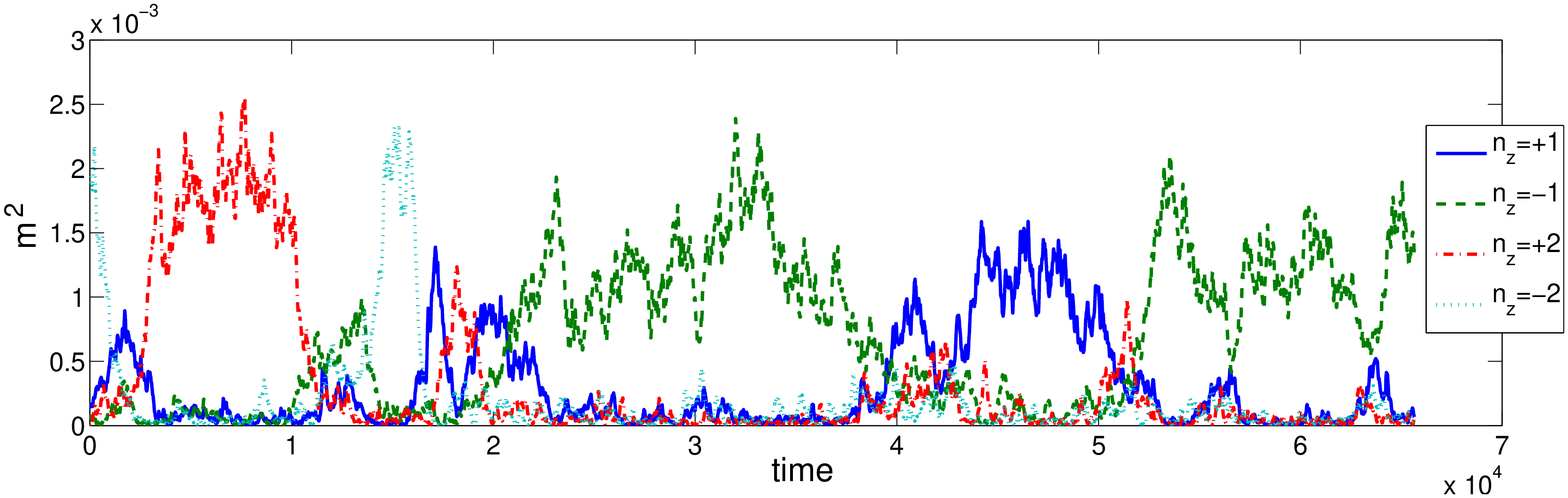}
\EC
\caption{Time series of $m^2(t)$ for several wave numbers $n_z=\pm1, \pm2$
for $R=315$ in a system of size $L_x\times L_z=128\times84$.\label{fig3}}
\EF
A pure pattern stage corresponds to a single $m(n_x,n_z)$ fluctuating
around a non zero value, the other $m(n_x',n_z')$ remaining negligible.
For instance the pattern keeps wave-number $n_z=+2$ from $t=3\,10^3$ to
$t=10^4$. The defective stage corresponds to $m(n_x,n_z)$ decaying
to zero while another one $m(n_x',n_z')$ grows. Wavenumbers $(n_x',n_z')$
may be different from $(n_x,n_z)$, in which case there is an effective
change of the orientation if $|n_z'|=|n_z|$ or a change of wavelength (sometimes combined with orientation changes) if $|n_z'|\ne|n_z|$.
In Fig.~\ref{fig3}, a change of orientation takes place at time
$t=4\,10^4$ ($n_z=-1\to+1$), a change of wavelength at time
$t=1.7\,10^4$
($n_z=-2\to+1$), the pattern with $n_z=+1$ growing back from a
defective stage at time $t=4.3\,10^4$.
Most of the time there is no ambiguity about the value of $n$ involved so that we shall use simplified notations, i.e. just
$m$ or $m_\pm$ instead of $m(n_x,\pm n_z)$, as often as possible.

Except very close to \RT,
pure state intermissions last long and defective episodes are short, so
that series of lifetime $T_i$ of well-oriented lapses can be defined from
recording simulations of duration sufficient to make reliable statistics.

\subsection{Lifetime computations\label{slc}}

Orientation and wavelength fluctuations are best characterised by
lifetimes distributions. Beforehand, we have to define
a systematic method to detect the beginning and the end of pure
pattern episodes from the $m^2$ time series. This is done by using
two thresholds: one, $s_1$, for the start
of a pure pattern episode and the other, $s_2$, for its termination,
see Fig.~\ref{fig4} (top).
\BF
\BC
\includegraphics[width=0.9\TW,height=0.3\TW,clip]{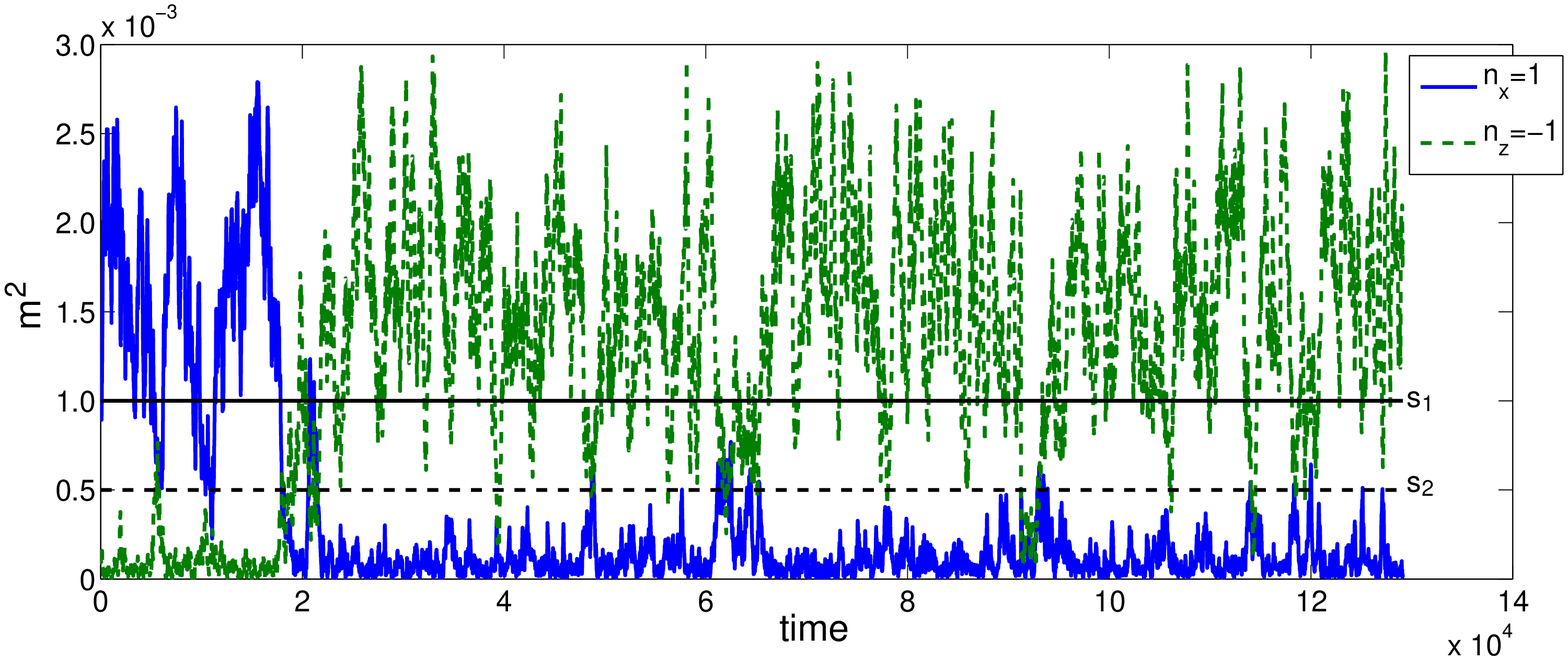}\\
\includegraphics[width=0.9\TW,height=0.3\TW,clip]{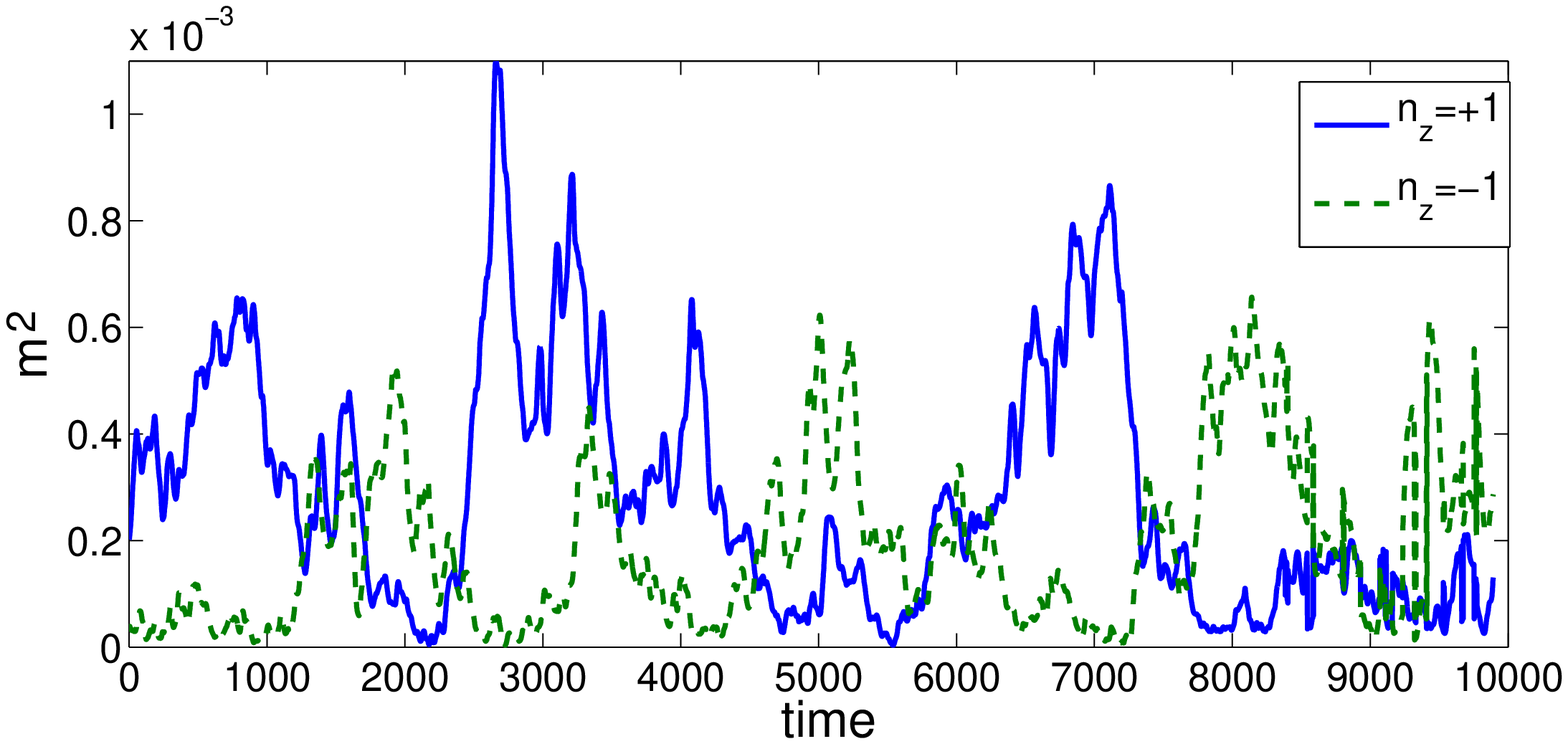}
\EC
\caption{Time series of $m^2$ for $n_z=\pm1$ at $R=330$ (top) and $R=345$ (bottom).
The two horizontal lines in the top panel locate threshold $s_1$ (full line) and $s_2$ (dashed line). Close to $R_{\rm t}$, bottom panel, orientation fluctuations are short-lived and much smaller, rendering the detection of well-oriented episodes more difficult.\label{fig4}}
\EF
The fast growth of $m^2$ makes it easy to choose $s_1$ and the results
are not much sensitive to its exact value.
In contrast, detecting the decay is
more problematic. This will be discussed in detail after the presentation
of a typical result obtained by assuming that the difficulty has been
properly resolved.

For practical reasons, we use a byproduct of the cumulated probability density
function (PDF) $Q$:
\BM
Q(T)=\frac{\#\{T_i\ge T \}}{\# \{ T_i\}}=1-\frac{\#\{T_i\le T \}}{\# \{ T_i\}}\,.
\EM
Empirical distributions obtained in an experiment with $L_x=110$ and
$L_z=32$ for $R=330$ are displayed in Fig.~\ref{fig5}. They
\BF
\BC
\includegraphics[height=7cm]{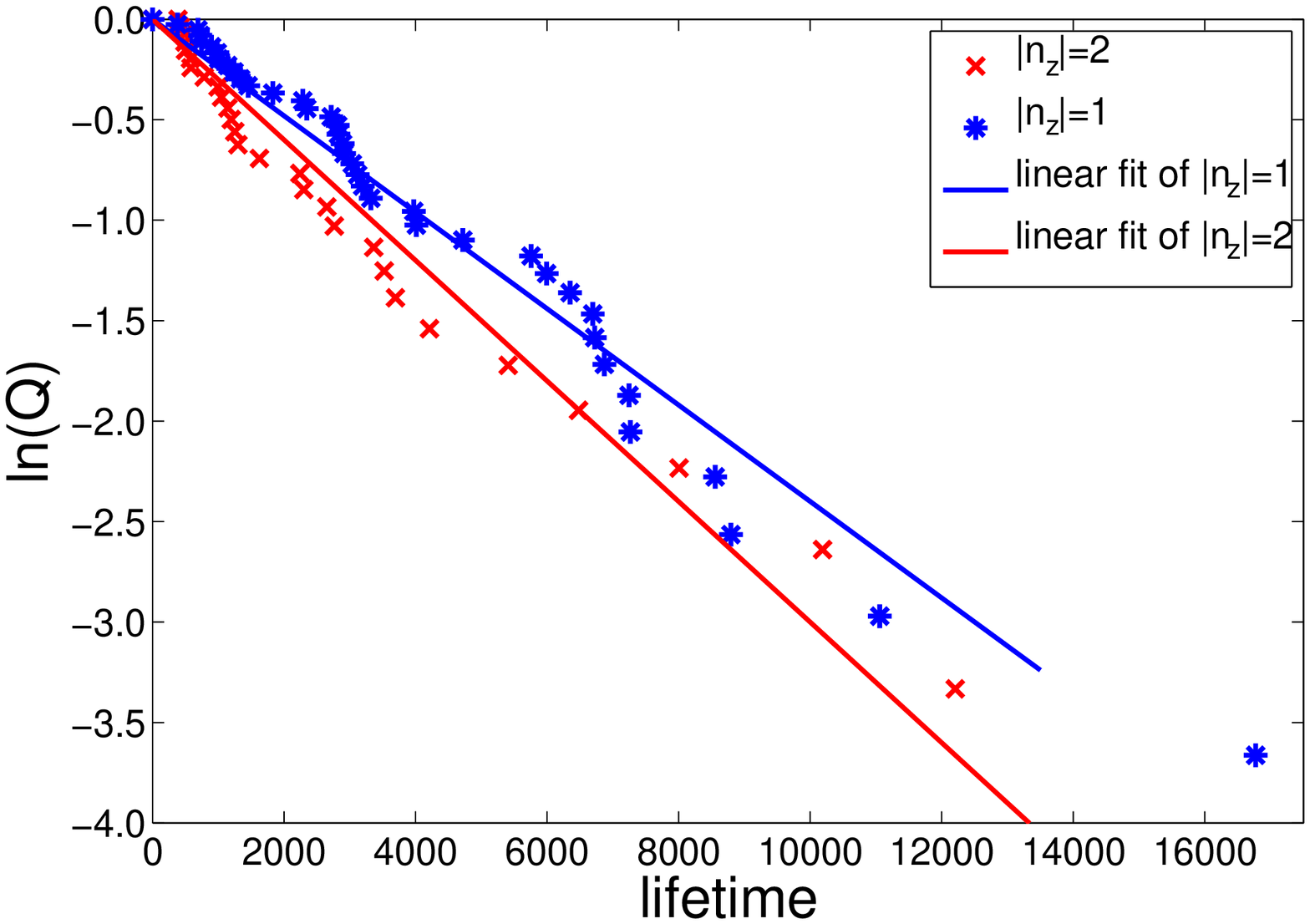}
\EC
\caption{Logarithm of $Q$ (right) for $L_x\times
L_z=128\times  84$, $R=315$, computed with $s_1=0.001$,
$s_2=0.005$, for both wave numbers $|n_z|=1$ and $|n_z|=2$. We have about 40 events for $|n_z|=1$ and about 20 events
for $|n_z|=2$. \label{fig5}}
\EF
are obtained from the time series, a small part of which is shown in
Fig.~\ref{fig4}, distinguishing $n_z=\pm1$ from $n_z=\pm2$. Since, for
symmetry reasons, the two orientations are supposed to have identical distributions, we sum over the $\pm$ in each case.
The semi-logarithmic coordinates used to represent $Q(T)$ suggest
exponentially decreasing variations, which makes orientation changes
look like deriving from a Poisson process. Assuming that they are
indeed in the form $\exp(-T/\langle T\rangle)$, we can obtain the mean
lifetime $\langle T\rangle$ from the plain arithmetic average of
the liftetime series.  $\langle T\rangle$ can also be obtained by fitting
the empirical cumulated distribution against an exponential law
or its logarithm against a linear law. In addition to raw data,
Fig.~\ref{fig5} displays the second kind of fits for $|n_z|=1$ and~$2$.
These three different estimates are close to each other
provided that the lifetime series comprise sufficiently large numbers of
events. An average of these three values will be used to define the mean
lifetime and the corresponding unbiased standard deviation will give
an estimate of the ``error'' for each lifetime series. 

Let us now come to the problem of the sensitivity of $Q(T)$ to the
value of the thresholds $s_1$ and $s_2$ used to determine the lifetimes
of the pure pattern episodes.
In Figure~\ref{fig6} (left) the mean lifetime $\langle T\rangle$ displays
a clear plateau as a functions of $s_1$. The width of this plateau does
not depend on $s_2$ though its value depends on it.
\BF
\BC
\includegraphics[width=0.49\TW,height=0.375\TW,clip]{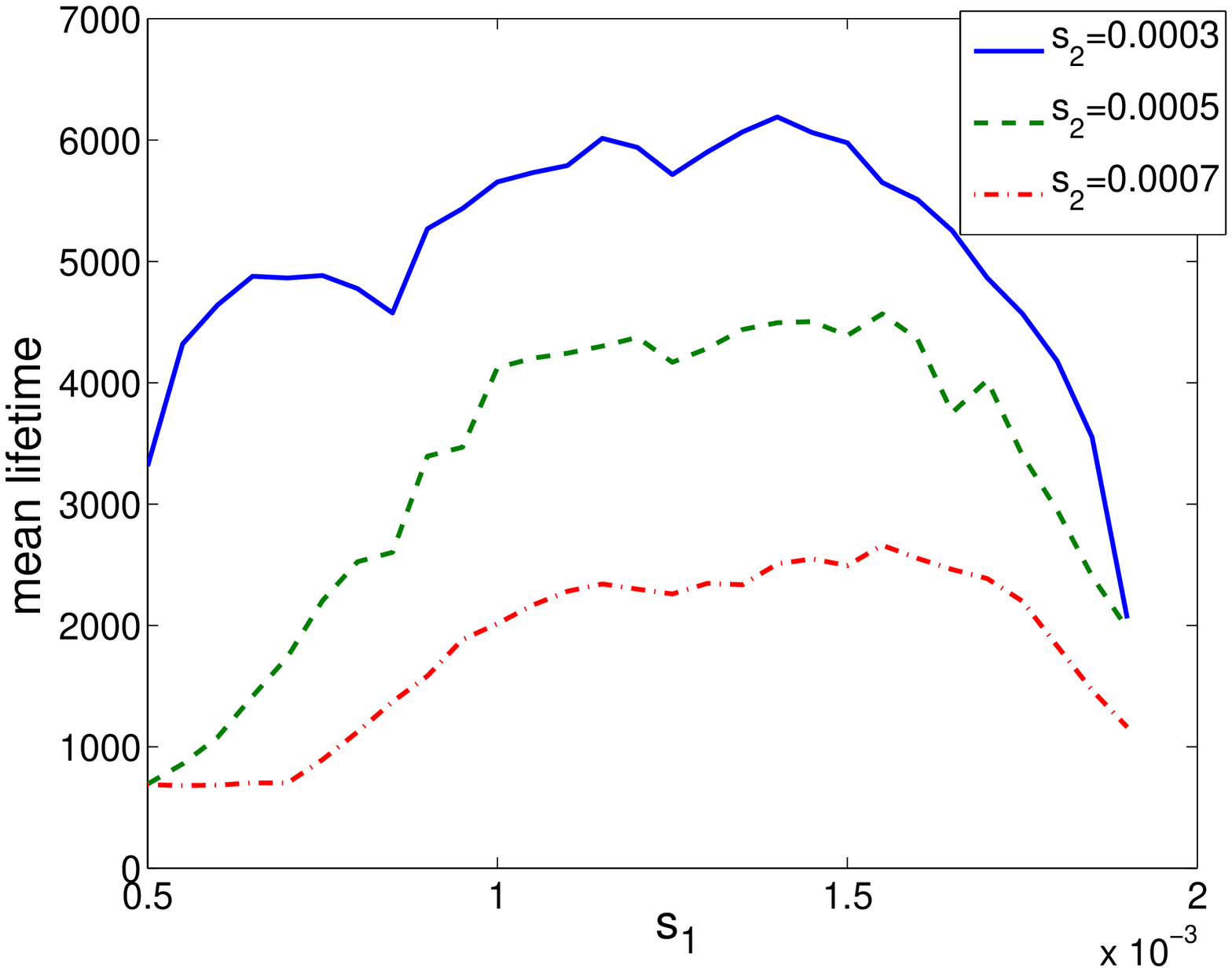}
\includegraphics[width=0.49\TW,height=0.375\TW,clip]{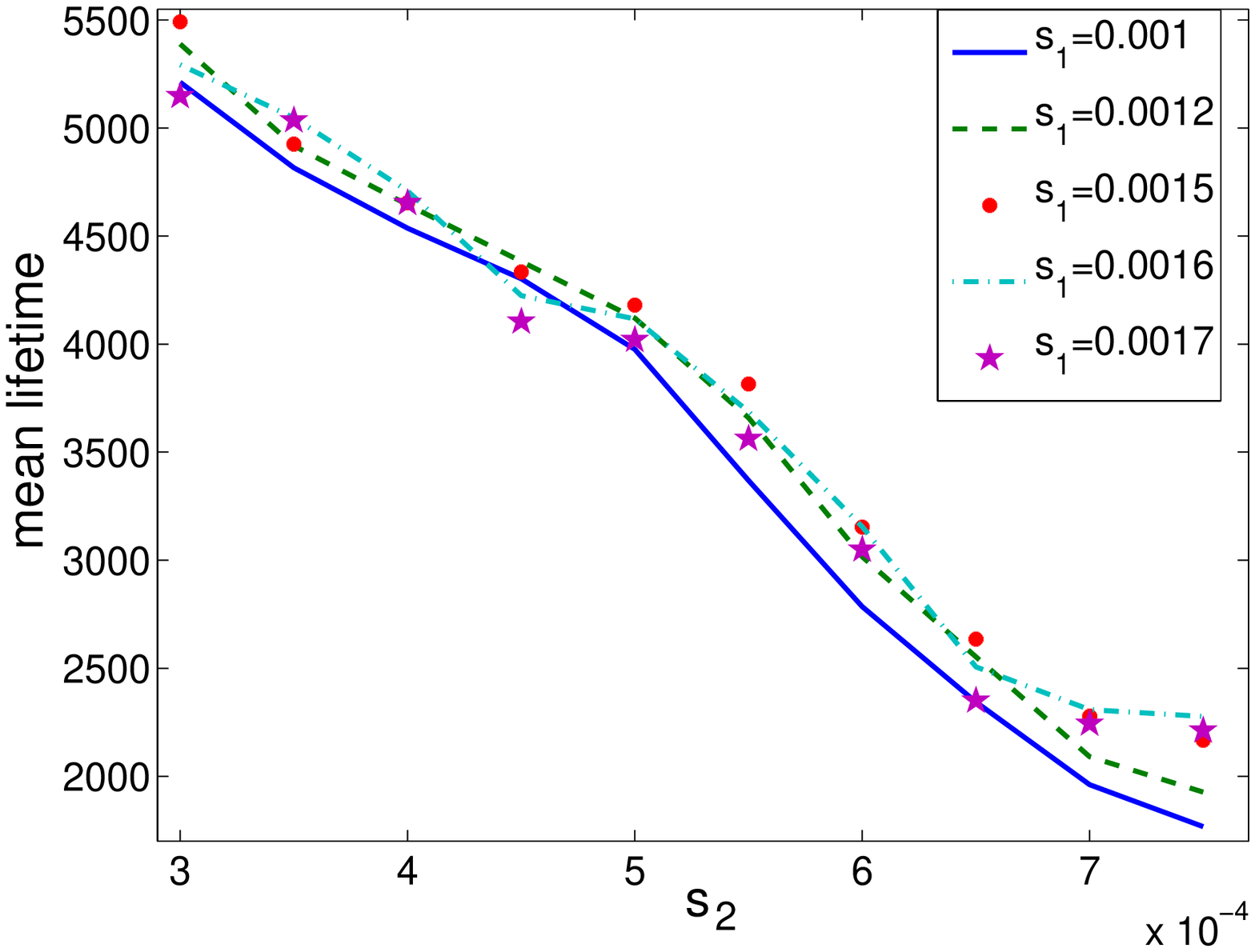}
\EC
\caption{Mean lifetimes functions of $s_1$ given $s_2$ (left) and
of $s_2$ given $s_1$ (right).  $R=315$ and
system size $L_x\times L_z=128\times 84$, $|n_z|=1$.} \label{fig6}
\EF
The existence of this plateau is easily seen to be related to the fast
growth of $m$ when the pattern sets in: $m$ always goes through
most of the values corresponding to the plateau in a very short time.
In practice, for $10^{-3}\le s_1\le 1.5\,10^{-3}$ the very same episodes
are detected whatever the precise value of $s_1$. That the plateau value still depends on $s_2$ just expresses that the duration of the
detected episodes are modified in the same way due to changes in the
detection of their termination. Of course, when $s_1$ is taken too large,
some less-well ordered episodes escape detection or are detected too late,
which artificially decreases the mean. On the other hand, if $s_1$ is
taken too small, the ``signal'' gets lost in the ``noise'': a large
number of brief noisy excursions are detected as relevant ordered episodes, again decreasing the mean.

The variation of the mean lifetime with $s_2$ is completely different
as seen in Fig.~\ref{fig6} (right). Here, $\langle T\rangle$ varies roughly
linearly with $s_2$ in a wide interval above the noise level
($\sim 3\,10^{-4}$, see Fig.~\ref{fig4}):
\BM
\langle T\rangle(s_2)\simeq a(1-b s_2)\,.
\EM
Coefficient $b=1000\pm100$ does not vary significantly over the cases that
we have considered. This dependence fully explains the change of plateau
value in plots of $\langle T\rangle$ as a function of
$s_1$. Coefficient $a$, corresponding to $\langle T\rangle$ extrapolated
toward $s_2=0$ however still depend on $R$ and the geometry. Henceforth,
we define this extrapolated value as the relevant average lifetime
$\langle T\rangle$, which will be supported by the theoretical
considerations to be developed in the next section.

The observed dependence of $\langle T\rangle$ on $s_2$ can be explained
by the fact that the decay of a pure pattern is much more gradual
than its growth, which causes significant differences when the
duration of an episode is measured, leading to a decrease of
$\langle T\rangle$ as $s_2$ increases since the termination of
the episode is detected earlier. A second reason why the mean lifetime
increases as $s_2$ decreases arises from the fact that some excursions
are not counted as decay events. In physical space, this corresponds
to an irregular and slow disorganisation of turbulence, contrasting
with the fast installation of the pattern. In fact  $\langle T\rangle$
cannot be obtained otherwise than by extrapolation of threshold
$s_2$ to zero, as will be discussed in \S\ref{ms}.

\subsection{DNS results\label{s4}}

The two systems sizes, $L_x\times L_z=128\times 64$ and $110\times 32$,
already considered in our previous work \cite{MRtcfd,RMepjb} are studied
here over the whole range of Reynolds numbers where the pattern exists
at the chosen numerical resolution, $R\in[\RG,\,\RT]=[275,345]$.
Orientation fluctuations are systematically found close enough to \RT, see Cases 1 \& 2 below.

In addition, wavelength fluctuations can take place when the size of the
system is too far away from resonating with the pattern's elementary
cell $\lambda_x^{\rm opt}\times\lambda_z^{\rm opt}$, where `opt' means
`optimal', in a sense to be defined below in \S\ref{llm}. Orientation {\it and\/}
wavelength fluctuations are observed at $R=315$
for $L_x=128$, $L_z=84$ and $90$, and for $L_x=110$, $L_z=84$, meaning
that both $|n_z|=1$ and  $|n_z|=2$ are competitive for $L_z=84$ or
$L_z=90$. In contrast,
lifetimes of single mode patterns are extremely long for $L_z<84$ and
$L_z>90$, meaning that $L_z<84$ is optimal for $|n_z|=1$ and
$L_z>90$ is optimal pour $|n_z|=2$. Orientation and wavelength fluctuations are similarly present
in several other circumstances, at
lower Reynolds number $R=272$ and $R=275$ for $L_x\times L_z=110\times 32$, 
as well as at $R=290$ for $L_z=48$ and $L_x=80$ or at $R=330$ for $L_x=90$, 
$140$, and $150$.

\paragraph{Case 1: $L_x=128$, $L_z=84$, $R=315$, wavelength
fluctuations.} Several experiments under the same protocol have been performed, using different initial conditions. Integration
times ranged from $5\,10^4$ to $10^5$~$h/U$. A large enough ensemble of
lifetimes has been sampled, both for $|n_z|=1$ and $|n_z|=2$, allowing us to compute the corresponding order parameters
$M$ -- the conditional time averages of $m(t)$ as defined in (\ref{E-m}) --
with sufficient accuracy. Snapshots corresponding to this aspect ratio
are displayed in Fig.~\ref{fig1}, a typical part of the
corresponding time series is shown in Fig.~\ref{fig3}.
For $|n_z|=1$ and $|n_z|=2$, we obtain $M_1=0.033\pm0.001$
and $M_2=0.038\pm0.001$, respectively. From the lifetime distributions
in Fig~\ref{fig5}, we get $\tau_1=8100\pm 200$  and $\tau_2=3800\pm100$.
 The fact that $M_1<M_2$ is not surprising and is understood in term of optimal wavelength (\S\ref{llm}, $\lambda^{\rm opt}_z \simeq 39$ at $R=315$ \cite{RMepjb}). The reason why one has $\tau_1>\tau_2$ is however not
clear.

\paragraph{Case 2: $L_x=110$, $L_z=32$, variable $R$, orientation
fluctuations.} A thorough account of the behaviour of $M$ and the re-entrance featureless
turbulence has been given in \cite{RMepjb}. Here, lifetimes are computed
for Reynolds number ranging from $R=325$ to $R=340$. Below $R=325$, the lifetimes are so long that a small number of events is observed despite
the length of time series used ($>2\,10^5$), which forbids the
determination of $\tau$ as a meaningful average (Fig. \ref{fig10}).
Above $R=340$, a clear separation of time scales is lacking, which now
forbids the definition of lifetimes of individual events, compare the two panels in Fig.~\ref{fig4}.

Figure~\ref{fig11} displays the variation of the average lifetime $\tau$
with $R$, showing that it increases by a factor of 10 as $R$ decreases
from $R=340$, which is somewhat below $\RT=355$, down to $R=325$ below
which it is too long to be measured reliably. ``Error bars'' suggested by
up and down triangles in Fig.~\ref{fig11} correspond to the unbiased
standard deviation of the three estimates for $\tau$ mentioned earlier.
\BF
\BC
\includegraphics[height=6cm]{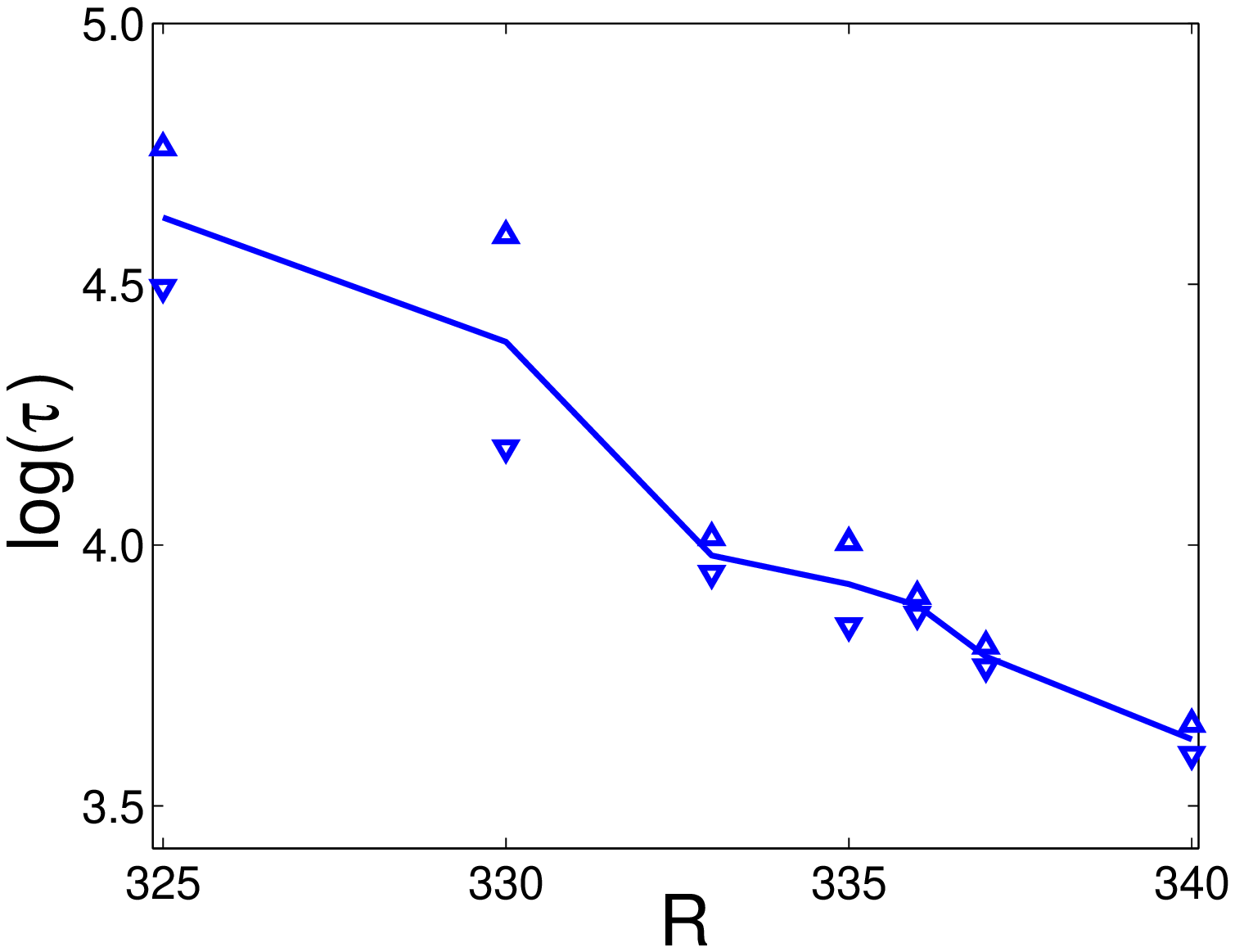}
\EC
\caption{Mean lifetime $\tau$ as a function of $R$ for
$L_x\times L_z=110\times 32$ (log scale).} \label{fig11}
\EF

\section{Conceptual framework and application to DNS results\label{s3}}

\subsection{The Landau--Langevin model\label{llm}}

Prigent {\it et al.} proposed to consider the turbulent bands
as resulting from a conventional pattern formation problem described at lowest order, from symmetry arguments, by
two coupled cubic Ginzburg--Landau equations, one for each band orientation, further subjected to noise featuring the turbulent background above $R_{\rm t}$. The slowly varying part of the velocity field component away from the laminar profile can be written as \[u_x=A_+(\tilde x,\tilde z,\tilde t)e^{ik_x^{\rm c}x+ik_z^{\rm c}z}+A_-(\tilde x,\tilde z,\tilde t)e^{ik_x^{\rm c}x-k_z^{\rm c}z}+cc\,,\] where $A_\pm \in \mathbb{C}$ are the amplitude fields accounting for
the two modulation waves, and $\tilde x$, $\tilde z$ and $\tilde t$ are slow variables \cite{Ma10}. Then, following this approach, we assume
\BE
\tau_0\partial_{\tilde t} A_\pm=
(\epsilon+\xi_x^2\partial_{\tilde x\tilde x}^2
+\xi_z^2\partial_{\tilde z\tilde z}^2)
A_\pm - g_1|A_\pm|^2A_\pm-g_2|A_\mp|^2A_\pm + \alpha\zeta_\pm\,,
\label{E-rnd}
\EE
the quantity $\epsilon=(R_{\rm t}-R)/R_{\rm t}$ measures
the relative distance to the threshold%
\footnote{The existence of a well defined threshold in this system is attested by the behaviour of the turbulent fraction and spatially averaged kinetic energy which display a marked change of slope at $R_{\rm t}$\cite{RMepjb}} \RT, $\tau_0$ is the `natural' time
scale for pattern formation, $\xi_{x,z}$ are streamwise and
spanwise correlation lengths, $g_1$ and $g_2$ are the self-coupling
and cross-coupling nonlinear coefficients, and $\alpha$ the strength
of the noise $\zeta_\pm$ supposed to be a centred Gaussian process with
unit variance. The strength $\alpha$ of the noise is expected to grow
smoothly with $R$, regardless of the existence of the pattern since the local intensity of the turbulence is empirically not directly correlated to the amplitude and phase of the modulation $A_\pm$.
The tilde variables describe the long-wave modulations to an ideal pattern
with critical wavelengths $\lambda_{x,z}^{\rm c}$ to which correspond
critical wavevectors $k_{x,z}^{\rm c}=2\pi/\lambda_{x,z}^{\rm c}$, the term critical referring to the most unstable wave vector near $R=R_{\rm t}$. The
systems that we consider have periodic boundary conditions placed at
distances $L_{x,z}$. Fourier analysis then leads to characterise the
pattern by
wavevectors ${\bf k}=(k_x,k_z)$, with $k_{x,z}=2\pi n_{x,z}/L_{x,z}$.
It is assumed that the wave numbers obtained during a given
experiment will be the integers that will be as close as possible of
 $n_{x,z}^{\rm c}=L_{x,z}/\lambda_{x,z}^{\rm c}$.
Furthermore, our systems can accommodate a small number of cells of
size $(\lambda_x,\lambda_z)$ so their modes are well isolated
\cite[Ch.4]{Ma10}.
Assuming that a single pair $(n_x,\pm n_z)$ is involved, the partial differential equation (\ref{E-rnd}) is turned into an ordinary differential
equation for $A(n_x,\pm n_z)$ simply denoted $A_\pm\equiv A_\pm^{\rm r}+i A_\pm^{\rm i}$,
close enough to \RT \cite{RMepjb}:
\BE
\tau_0\DDt A_\pm
=\tilde\epsilon A_\pm-g_1|A_\pm|^2A_\pm-g_2|A_\mp|^2A_\pm+
\alpha\zeta_\pm\,,\label{lan}
\EE
where $\tilde\epsilon=\epsilon-\xi_x^2 \delta k_x^2-\xi_z^2\delta k_z^2$
controlling the linear stability of these modes, is evaluated for
$\delta k_{x,z}= k_{x,z}-k_{x,z}^{\rm c}$ with the relevant
$k_{x,z}=2\pi n_{x,z}/L_{x,z}$, as well as the nonlinear coefficients
$g_{1,2}$ ($\in\mathbb{R}$ because the pattern does not drift, at least in the absence of noise). Coefficient $\alpha$ is the effective
strength of the noise affecting the mode that we consider.
Equation (\ref{lan}) can be written as deriving from a potential:
\BM
\tau_0\DDt A_\pm^{\rm r,i}=
-\frac{\partial\mathcal{V}} {\partial A_\pm^{\rm r,i}}+
\alpha \zeta_\pm\,,
\EM
with
\BE
\mathcal{V}=
-\sfrac12\tilde\epsilon\left(|A_+|^2+|A_-|^2 \right)
+\sfrac14 g_1\left(|A_+|^4+|A_-|^4 \right)
+\sfrac12 g_2 |A_+|^2|A_-|^2\,. \label{pot}
\EE
Usually, when making use of phenomenological equations such as (2), one relies on values of critical wavevectors $k^{\rm c}$ that are computed once for all from some linear stability theory and further introduced in the perturbation expansions solving the nonlinear wavelength selection problem   beyond the threshold \cite{CrHo93}. Here the theory is not developed enough  to have such a definition and such an evaluation of nonlinearly selected  `optimal' wavevectors far enough from threshold. Accordingly, in (3) we introduce values of $\tilde\epsilon$ that do not make reference to some explicit computation involving measured values of $\epsilon$ and $\xi_{x,z}$ but values that are just estimates consistent with the empirically determined optimal wavelengths. In the same way, we keep the cubic Landau expressions (3), neglecting higher order terms that would introduce too many little-constrained parameters, without deeper insight into the problem.

The stable fixed points of the deterministic part of (\ref{lan})
were shown to correspond to the permanent state of the pattern
and the additive noise term seen to account for fluctuations
quite well by solving the corresponding Fokker--Planck equation
\cite{RMepjb}. The stationary probability distribution for the moduli
$|A_\pm|=A_\pm^{\rm m}$ was obtained in the form:
\BE
\Pi(A_+^{\rm m},A_-^{\rm m})
= Z^{-1}
A_+^{\rm m}A_-^{\rm m}\exp(-2\mathcal V/\alpha^2)\,,\qquad
Z=\int A_+^{\rm m}A_-^{\rm m}\exp(-2\mathcal V/\alpha^2)
\,{\rm d}A_+^{\rm m}{\rm d}A_-^{\rm m}\,.
\label{pdf}
\EE
The time behaviour of $A_\pm^{\rm m}$ is easily discussed by
considering the shape of $\mathcal V$ within the stochastic process
framework. Two limiting cases can be identified, depending on whether
$\tilde\epsilon$ is $\mathcal O(1)$ or $\ll1$. In the first case, excursions from the neighbourhood of the minima of $\mathcal V$ are
rare; the lifetime of an ordered episode can be defined as the average
time necessary for the system to go from the neighbourhood of a minimum
to the potential's saddle. It is expected to increase with the height of
the potential barrier, i.e. as parameter $\tilde\epsilon$ grows, and to
fall off as $\alpha$ increases. The lack of symmetry between the
growth and the decay of a pattern has then a clear explanation when
$\tilde\epsilon$
is large: The growth corresponds to the  system falling from the neighbourhood of the saddle into one of the wells; even in the presence of noise, this evolution is fast and mostly deterministic. In contrast, the
decay corresponds to the system slowly climbing toward the saddle against
the deterministic flow, driven by the sole effect of noise. In the
opposite limit, when $\tilde\epsilon$ approaches zero, the definition of a
lifetime no longer makes sense since the characteristic times for growth
and decay become of the same order of magnitude. 

\subsection{Orientation lifetimes from the model\label{ltt}}

Orientation changes and associated lifetimes are analysed in terms
of {\it first passage time\/} and {\it escape from metastable states\/}   \cite{VK90}. The distribution of lifetimes is anticipated to be Poissonian
as expected from a jump process controlled by an activation
``energy''. In a simplified one-dimensional version of potential $\cal V$
\cite[Ch.11,\S2,6--7]{VK90}, if the well is deep enough, within a parabolic
approximation the mean escape time, the average time necessary to go from a well to another, is given by
\BE
\tau/\tau_0=
\frac{2\pi}{\sqrt{\mathcal{V}_{\rm w}''\,|\mathcal{V}_{\rm s}''|}}
\exp\left(2
\frac{\mathcal{V}_{\rm s}-\mathcal{V}_{\rm w}}{\alpha^2}\right)\,,
\label{tau}
\EE
where `w' stands for `well' and `s' for `saddle';
$\mathcal{V}_{\rm w,s}$ are the values of the potentials at the
corresponding points and $\mathcal{V}_{\rm w,s}''$ the values of the
second order derivatives of the potential with respect to the variable at these
points. The derivation of this formula shows that $\tau$ is dominated by the time spent around the saddle.
 In our two dimensional system with potential (\ref{pot}),
at lowest order in $\alpha$ the coordinates of the well and saddle
points are:
\BM
(A_\pm^{\rm w},A_\mp^{\rm w})=\left(\sqrt{\tilde\epsilon/g_1},\,0\right)
\qquad \mbox{and} \qquad
(A_+^{\rm s},A_-^{\rm s})=\left(\sqrt{\tilde\epsilon/(g_1+g_2)},\,
\sqrt{\epsilon/(g_1+g_2)}\right)
\EM
and the corresponding values of the potential:
\BM
\mathcal V_{\rm w}=-\tilde\epsilon^2/4g_1\qquad\mbox{and}\qquad
\mathcal V_{\rm s}=-\tilde\epsilon^2/2(g_1+g_2)\,.
\EM
The second derivatives
have to be replaced by the eigenvalues of the Hessian matrix of $\cal V$
computed at these points:
\BM
H_{\rm w}=
\left(\begin{matrix}
2\tilde\epsilon&0\\
0&\tilde\epsilon(g_2/g_1-1)
\end{matrix}\right)
\qquad\mbox{and}\qquad
H_{\rm s}=
\frac{2\tilde\epsilon}{g_2+g_1} 
\left(
\begin{matrix} g_1&g_2\\g_2&g_1 \end{matrix}
\right)\,.
\EM
At point `w', $H_{\rm w}$ is diagonal and the eigen-direction
pointing to point `s' has eigen-value $\tilde\epsilon(g_2/g_1-1)$.
At point `s', $H_{\rm s}$ is diagonal in the basis $\{(1,1),(1,-1)\}$
and has eigenvalues
$\{2\tilde\epsilon,\,2\tilde\epsilon(g_1-g_2)/(g_1+g_2)\}$. The unstable
eigen-direction correspond to the second one
which is negative ($g_2>g_1$). Inserting these values in (\ref{tau}),
we obtain:
\BE
\tau/\tau_0=\frac{2\pi\sqrt2}{\tilde\epsilon}
\frac{\sqrt{g_1(g_1+g_2)}}{g_2-g_1}
\exp\left(\frac{\tilde\epsilon^2}{2\alpha^2g_1}\,
\frac{g_2-g_1}{g_1+g_2}\right)\,.\label{ttau}
\EE
In its exponential factor, this formula points out an ``energy'' scale
$\tilde\epsilon^2/g_1$ to be compared to the characteristic noise
energy $\alpha^2$ which play the role of the Boltzmann energy in
thermal problems. It also shows that, especially when $g_2$ is larger
but comparable to $g_1$, the noise energy has to remain small enough
because the parabolic approximation which underlies the formula
assumes sufficiently deep wells. The main difference between the one and two dimensions cases are in the shape of the ``energy landscape'', corrections are therefore expected to be multiplicative and not depend on the value of $\tilde{\epsilon}$.

\subsection{Simulation of the model\label{ms}}

For the deterministic part of
equation (\ref{lan}) we use a simple first-order implicit Euler algorithm,
while the additive noise $\alpha\zeta(t)$ is treated as a Gaussian
random
variable with standard deviation $\alpha\sqrt{dt}$ at each time step.
The model is integrated over a range of $\tilde\epsilon$, given $g_1$,
$g_2$ (several values), and $\alpha$ (assumed constant). The time
series of $|A_\pm|^2$ displayed in Fig.~\ref{fig8} are indeed reminiscent
of those obtained by numerical integration of Navier--Stokes equations
(Figs. \ref{fig3} and \ref{fig4}): pure states at the bottom of the wells correspond to $|A_-|^2$ fluctuating around~0 and $|A_+|^2$ away from~0, or
the reverse.
\BF
\BC
\includegraphics[width=0.8\TW]{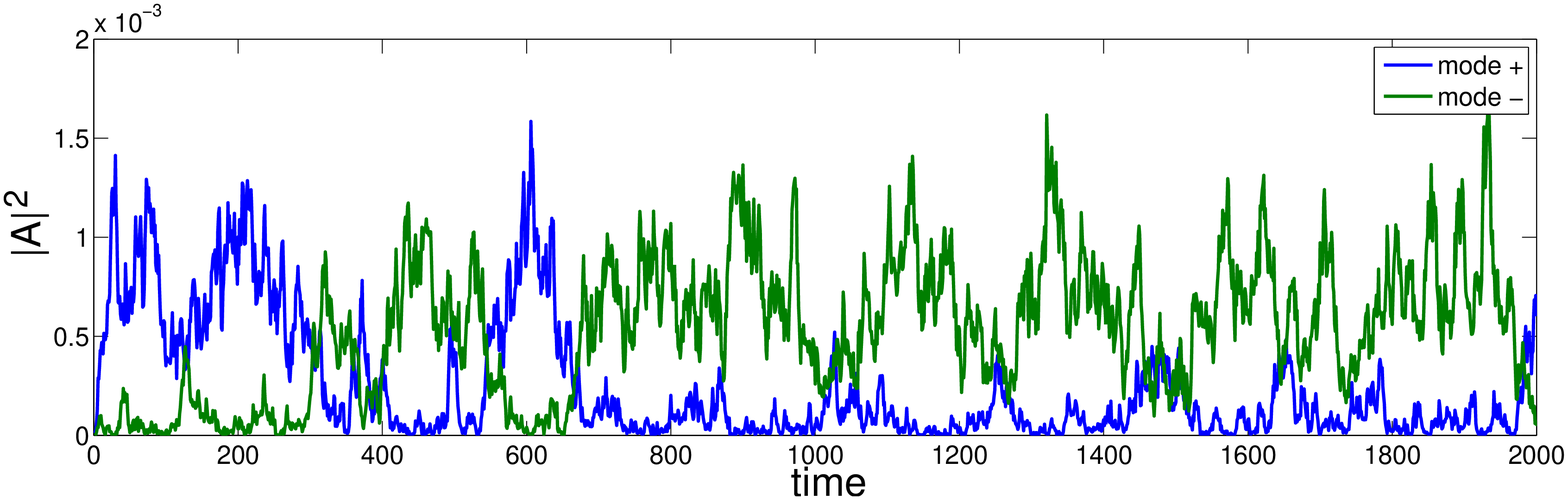}
\EC
\caption{Typical time series from model (\ref{lan}), for a set of
parameters corresponding to the Navier--Stokes DNS, $\tilde\epsilon=0.05$, $g_1=60$,
$g_2=120$, $\alpha=0.002$ \cite{RMepjb}.\label{fig8}}
\EF
Large excursions can lead to a change of the dominant orientation.
These excursions are more likely to occur when $\tilde\epsilon$ is decreased.

Lifetimes have been computed in the same way as for Fig.~\ref{fig6}.
The dependence of the mean lifetimes on thresholds~$s_{1,2}$ is displayed
in Figure~\ref{fig9}.
\BF
\BC
\includegraphics[height=6cm,clip]{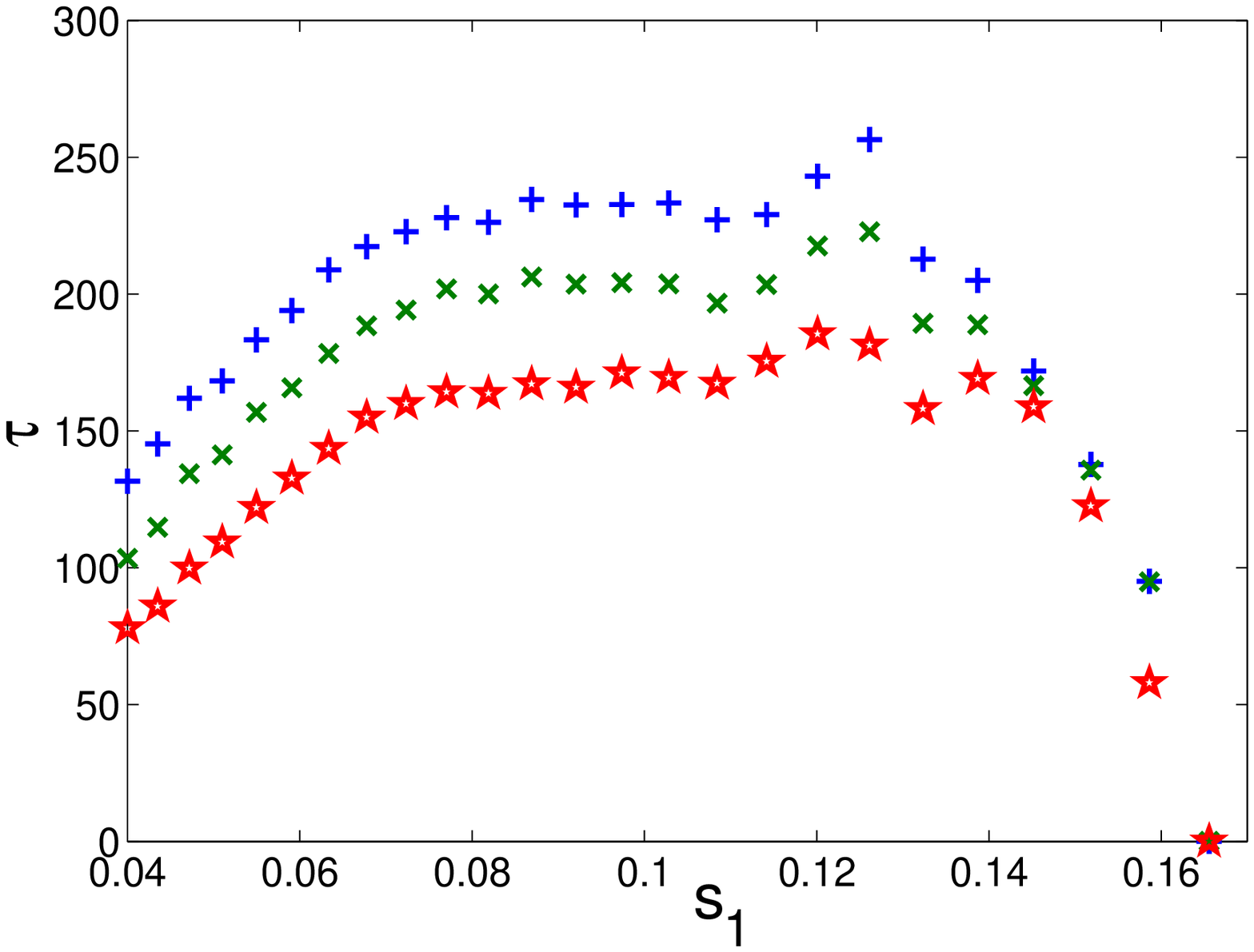}
\includegraphics[height=6cm,clip]{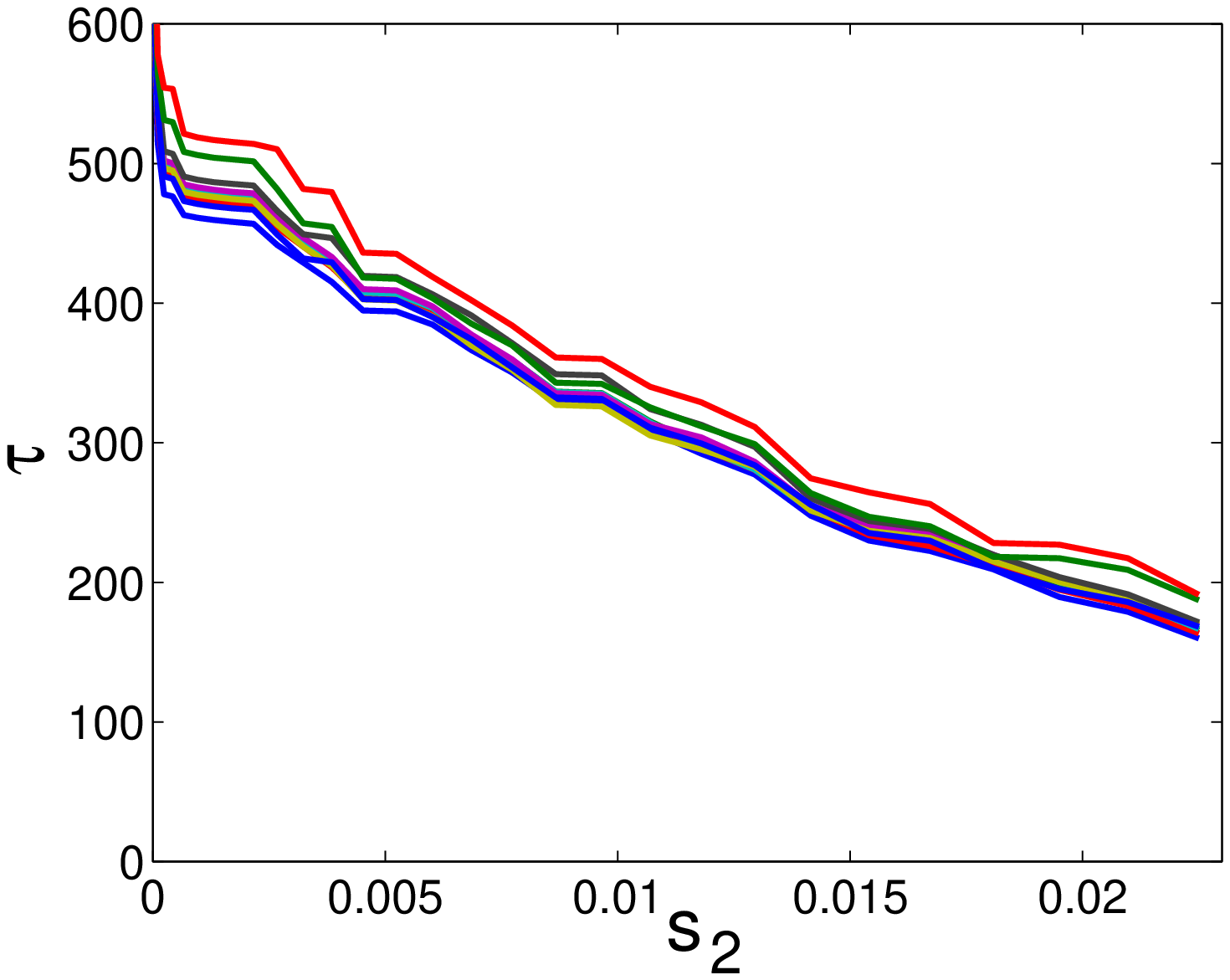}
\EC
\caption{Mean lifetime $\tau$, extracted from the model as a function of
$s_1$,
for $s_2=0.0163$, $0.0193$, and $0.0225$ (left) and as a function of $s_2$ for several $s_1$ ranging from $0.0784$ to $0.1296$ (right).
$\tilde\epsilon=0.075$, $g_1=1$, $g_2=2$, $\alpha=0.002$. }
\label{fig9}
\EF
A neat plateau is obtained for $s_1\in[0.075,0.115]$
for different values of $s_2$ (Fig.~\ref{fig9}, left), which corresponds
to the trajectory getting away from the saddle.
Extremes values of $s_1$ lead to bad estimates of $\tau$ for the same reasons as stated
before. As to threshold $s_2$, an orientation change has taken place when
the trajectory goes beyond the saddle, while a pure state
corresponds to one amplitude large and the other at the noise level.
We have thus to detect the change from large to small for one or the other
amplitude. It is extremely difficult to detect the precise passage at the saddle, since it is dominated by the time spent in that region, contribution from the two sides of the saddle point having the same weight. On the contrary the passage from one state to another leaves no doubt as to its definition. Therefore, we prefer to compute the mean first passage time from one well to another, which is obtained in our simulation by the extrapolation at $s_2=0$.
Approximating the curves in Fig.~\ref{fig9} (right) by linear functions
$\tau=a(1-bs_2)$, one finds that the slope $b$ depends on
$\tilde\epsilon$ and $s_1$ only weakly; the value of $\tau$ retained
is then the one given by the extrapolation $s_2=0$, i.e. coefficient $a$.
Improving the definition of $\tau$ with approximations better than the linear one has not been found necessary. The general expression of the mean first passage time gives no hint as to the quantitative behaviour on the distance to the second well $s_2$, although it shows the same qualitative behaviour as seen in Figures~\ref{fig6} (right) and~\ref{fig9} (right).
In Figure~\ref{fig10} (semilog coordinates), the lifetime $\tau$ measured
in this way is compared to the asymptotic expression from the theory
(\ref{ttau}) as a function of $\tilde\epsilon$. 
\BF
\BC
\includegraphics[height=6cm,clip]{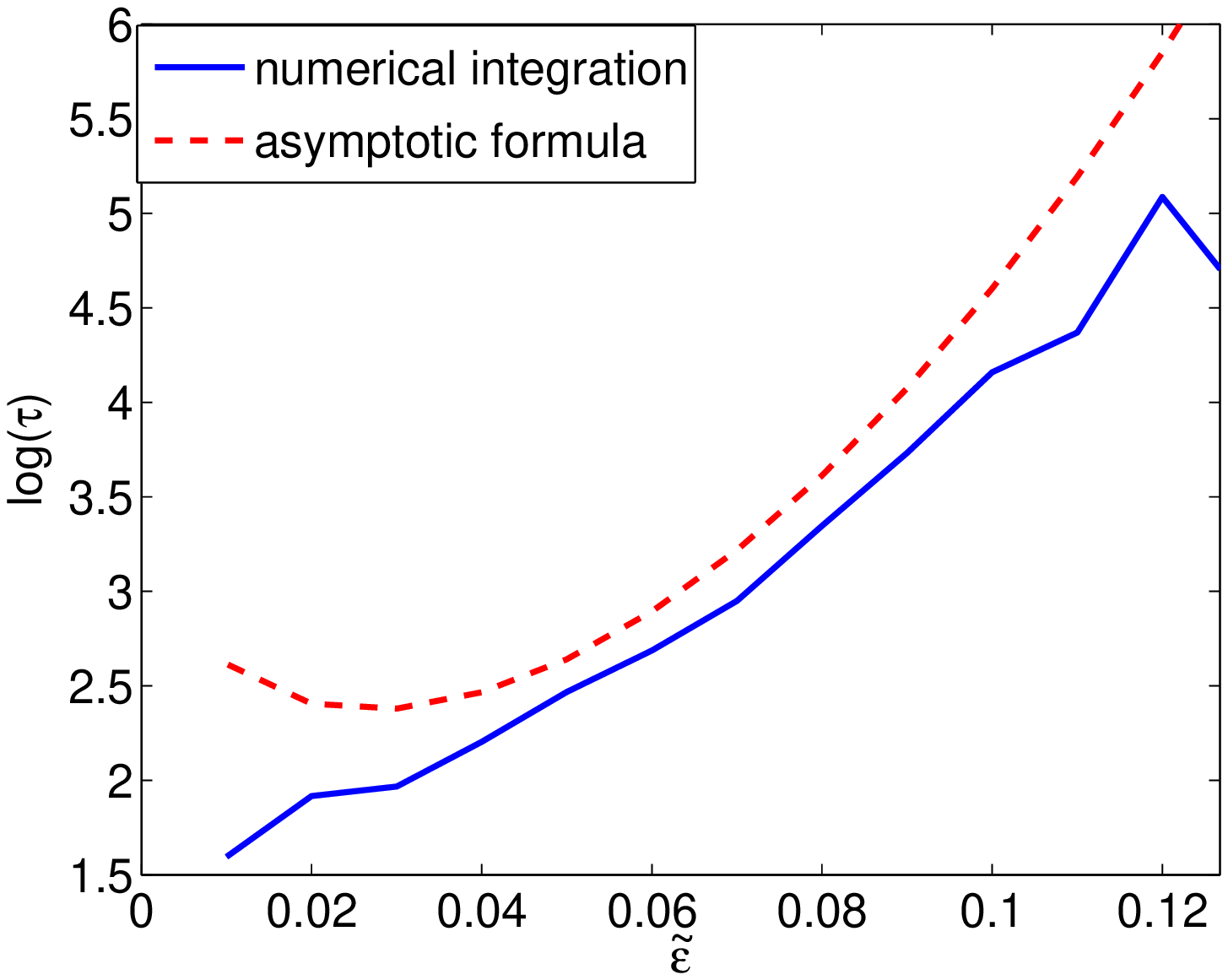}
\EC
\caption{$\log(\tau)$ as a function of $\tilde\epsilon$ for $g_1=60$,
$g_2=120$, $\tilde\alpha=0.002$; the model was integrated over $10^5$ time
units.\label{fig10}}
\EF
It can be seen that the asymptotic formula is not valid for the smallest
values of $\tilde\epsilon$, when the wells are not longer deep enough for the approximation to be valid, similarly to what is found in the DNS close to $R_{\rm t}$. The values given by this formula for small values of $\tilde{\epsilon}$, especially around and below the minimum it
predicts at $\tilde\epsilon=\alpha\sqrt{g_1(g_2+g_1)/(g_2-g_1)}$, cannot be trusted. For large
$\tilde\epsilon$, the lifetime computed from the simulation saturates
because it becomes of the order of the total integration time so that only
a few events smaller than this total time can be recorded. Accordingly the long time tail of the distribution is badly sampled with an under-representation
of lifetimes larger than the average expected from the theory. In Fig.~\ref{fig10}, the numerical and the asymptotic estimates of
the mean lifetimes are seen to differ by a constant of order unity,
which is attributed to the one-dimensional character of the approximation. 

\subsection{Generalisation}

This approach can be extended to wavelength fluctuations. When the size
$(L_x,L_z)$ of the system is such that it `hesitates' between two pairs
of modes $(n_x,\pm n_z)$ and $(n'_x,\pm n'_z)$, we introduce
two supplementary amplitudes $A(n'_x,\pm n'_z)$ that we denote $B_\pm$ for short and, extending notations straightforwardly with primes for quantities
related to $B_\pm$, we arrive at:
\BA
\tau_0\DDt A_\pm
=\tilde\epsilon A_\pm-g_1|A_\pm|^2A_\pm-g_2|A_\pm|^2A_\pm
+ g_3 (|B_\pm|^2+|B_\mp|^2) A_\pm+
\alpha\zeta_\pm\,,\label{lan1}\\
\tau_0\DDt B_\pm
=\tilde\epsilon' B_\pm-g'_1|B_\pm|^2B_\pm-g'_2|B_\pm|^2B_\pm
+ g'_3 (|A_\pm|^2+|A_\mp|^2) B_\pm+\alpha'\zeta'_\pm
\label{lan2}\,,
\EA
where $\tilde\epsilon$ and $\tilde\epsilon'$ as well as the nonlinear
coupling constants $g_{1,2,3}$, $g'_{1,2,3}$ and even the effective noise
intensities $\alpha$, $\alpha'$ may differ since they relate
to pure
patterns with different $\delta k_{x,z}=k_{x,z}-k_{x,z}^{\rm c}$. A first
guess would be to assume the primed and non-primed variables equal, which
would bring us immediately back to the previous approach with an effective
potential, wells, saddles, and potential barriers, leading to
estimates for the different lifetimes involved.

It is not clear how
the case of turbulence re-entrance (the intermittent regime of \cite{BT05})
would fit this framework but it is well described by a PDF with three
peaks  \cite{RMepjb} corresponding to a probability potential with three wells and thus hopefully amenable to a similar treatment with a similar
output.

These generalisations have not been worked out in detail numerically
since they introduces a discouragingly large number of parameters to
be fitted against the experiments and from which we would learn little,
owing to their phenomenological basis. Only the case involving a single
pair of modes was examined in \S\ref{ms} above, mostly
in order to validate the procedure followed to determine lifetimes
in \S\ref{slc}.

\section{Summary and conclusion\label{s5}}

In this paper, numerical simulations of the Navier--Stokes equation
in plane Couette flow configuration have been performed in a range of
Reynolds numbers where the transition to turbulence happens in the form
of oblique bands. Systems with sizes fitting a few elementary cells
$\lambda_x\times\lambda_z$ of the pattern have been considered.
These sizes are much larger than the minimal flow unit which allows
the reduction of the transition problem to a temporal process familiar
to chaos theory \cite{Eetal08}. Accordingly, the considered systems
are able to display the first manifestations of a genuinely
spatiotemporal dynamics {\it via\/} patterning. Following the patterns
in time, we showed that they experience orientation and wavelength
fluctuations in the upper part of the range of transitional Reynolds
numbers $[\RG,\RT]$. A systematic procedure to detect the start and the
termination of well-oriented episodes was defined, leading to the
observation of exponentially decreasing distributions for their
lifetimes (Fig.~\ref{fig5}).

A consistent interpretation scheme was then provided by adapting the noisy
Ginzburg--Landau model proposed in \cite{Petal03} to our case, transforming 
the original stochastic PDE into a Landau--Langevin stochastic ODE.
Besides supporting the procedure used to determine lifetimes, the approach
directly leads to the determination of probability distributions for
the patterned states from the shape of the potential obtained by solving
the corresponding Fokker--Planck equation, as already suggested
in~\cite[Fig.~19]{Petal03}. The variation of the patterns'
mean lifetimes is thus linked to the relative distance to threshold and
noise intensity through an asymptotic formula involving the ``energy''
barrier between wells corresponding to the different well-oriented states
in competition. Ingredients in the relative distance to threshold
$\tilde\epsilon$ which is a function of both the Reynolds number and
the optimal wavelength, are amply sufficient to explain most of the
dependance of the mean lifetimes as functions of $R$, $L_{x,z}$, and
the spontaneous appearance of defects separating patches of well-oriented
patterns close enough to \RT\ in larger aspect-ratio systems as illustrated in Fig.~\ref{fig13} and seen in the experiments \cite{Petal03}. 
\BF
\BC
\includegraphics[height=1cm,clip]{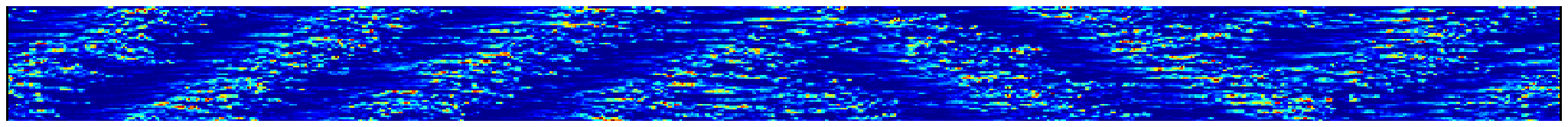}
\EC
\caption{Orientation defect spontaneously appearing in the flow for
$R=340$ in a domain of size $L_x\times L_z=660\times48$.\label{fig13}}
\EF

In order to explain the occurrence of exponentially decreasing lifetime
distributions, the theory of dynamical systems appeals to the sensitivity
to initial conditions of trajectories visiting a homoclinic
tangle~\cite{Eetal08}. Here, the modelling that fits well our observations
implies that exponential distributions arise from some jump random
process \cite{VK90}. As soon as the size of the system is much larger
that the minimal flow unit (for which the temporal behaviour inherent in
low dimensional dynamical systems is relevant), a spatiotemporal
perspective becomes in order, and the jumps in question can easily be
associated to the local chaotic dynamics of pieces of streaks and
streamwise vortices involved in the self sustaining process of turbulence \cite{Wa97}.
This local chaotic dynamics would then be responsible for the wandering of
the global system through some ``energy'' landscape with wells and saddles.
With system sizes of the order of the elementary pattern cell $\lambda_x\times\lambda_z$, this wandering amounts to orientation and/or wavelength
changes. Extending these views to larger systems would then explain
the statistical properties of fluctuating laminar-turbulent patches
observed in the upper transitional range close enough to \RT\
\cite{Petal03}.

Whereas the origin of the noise introduced in the
description is understandable from chaos at the local (microscopic)
scale, it remains however to understand why the coexistence of laminar
and turbulent flow takes the form of oblique bands at the global
(macroscopic) scale, i.e. to justify the Ginzburg--Landau approach
from the first principles rather than taking it as an educated
phenomenological guess.

\end{document}